\documentclass[12pt]{iopart}

\usepackage{amsmath}
\usepackage{setspace}
\usepackage{amssymb}
\usepackage{graphicx}

\usepackage{dcolumn}
\usepackage{bm}
\usepackage{hyperref}
\usepackage[cp1250]{inputenc}
\usepackage[usenames,dvipsnames]{color}
\allowdisplaybreaks[4]
\newcommand{\g}[1]{{\bf #1}}

\newcommand*{\he}{$^3$He}
\begin{document}

\title[]{Properties of an almost localized Fermi liquid in applied magnetic field revisited: Statistically consistent Gutzwiller approach}
\author{Marcin M. Wysoki\'nski}
\ead{marcin.wysokinski@uj.edu.pl}
\address{Marian Smoluchowski Institute of Physics$,$ Jagiellonian University$,$ Reymonta 4$,$ PL-30-059 Krak\'ow$,$ Poland}

\author{Jozef Spa\l ek}
\ead{ufspalek@if.uj.edu.pl}

\address{Marian Smoluchowski Institute of Physics$,$ Jagiellonian University$,$ Reymonta 4$,$ PL-30-059 Krak\'ow$,$ Poland}
\address{Academic Centre for Materials and Nanotechnology (ACMiN) and Faculty of Physics and Applied Computer Science, 
AGH University of Science and Technology, Al. Mickiewicza 30, PL-30-059 Krak\'ow, Poland}

\date{\today}

\begin{abstract}
We discuss the Hubbard model in an applied magnetic field and analyze the properties of neutral spin-$\frac{1}{2}$ fermions within the
so-called {\it statistically consistent Gutzwiller approximation} (SGA). The magnetization curve 
reproduces in a semiquantitative manner the experimental data for liquid \he\ in the regime of moderate correlations and in the presence 
of small number of vacant cells, modeled by a non-half filled-band situation, when a small number of vacancies ($\sim5\%$) is introduced
in the virtual fcc lattice. We also present the
results for the magnetic susceptibility and the specific heat, in which a {\it metamagnetic-like} behavior is also singled out in a non-half-filled 
band case. 
\end{abstract}

\maketitle
\section{Introduction}

In condensed matter physics the systems with moderate to strong local correlations 
such as almost-localized electron systems, heavy-fermion metals, 
liquid \he, and selected cold atomic systems in optical lattices, have been extensively 
studied during the last decades. On the other hand, starting from the seminal 
Landau papers on theory of Fermi liquid\cite{landau1959,landau1957} as 
a direct generalization of the concept of electron gas, it becomes unquestionable 
that interaction between fermions is the source of their nontrivial physical properties even
before the Mott transition to the localized state takes place for either sufficiently strong interactions or low density. 
Original Landau formulation\cite{landau1959}, despite being in essence phenomenological
provided, inter alia, a good qualitative rationalization of the observed effective mass enhancement 
of \he\ atoms in the liquid state\cite{greywall1983,greywall1986,mook1985,dobbs}
and that of electrons in metals. However, the 
Landau Fermi liquid theory turned out to be insufficient to account 
for specific more sophisticated effects of the correlations, such as the Mott (Mott-Hubbard) localization\cite{gebhardmott}
or the appearance of spin-dependent effective masses of quasiparticles\cite{gopalan1990}
or the observation of metamagnetism of itinerant almost localized and correlated fermions\cite{vollhardt1984}.
Note that by {\it correlated fermions} or a {\it correlated Fermi liquid} we understand the system for which the kinetic 
or Fermi energy (per particle) is comparable or even smaller than the interaction energy per particle. Therefore, 
strictly speaking, one should term those systems as {\it strongly interacting}, but that particular
phrase is reserved for high-energy interactions of elementary
particles. Similar correlated states appear also there, as e.g. the deconfinement transition
of a condensed hadron matter into the quark-gluon plasma contains the principal features of the 
Mott localization-delocalization transition\cite{castorina2010,satz2012}. Hence the concept 
of an almost localized Fermi liquid seems to have a universal meaning in condensed matter physics.

\begin{figure}[b]
\begin{center}
\includegraphics[width=80mm]{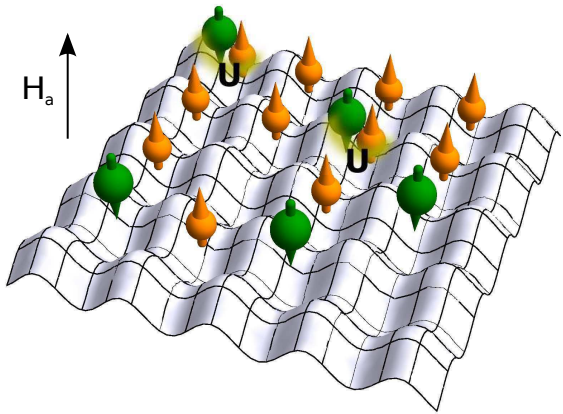}
\caption{Schematic visualization of division of quantum-liquid state into an effective 
lattice system composed of cells (modeled by potential valleys with the spin $\frac{1}{2}$ fermions).
Every cell can contain up to two fermions with opposite spins, at the price of energy loss $U$.
A number of fermions $N_e$ is assumed in general to differ from the number of all sites $N_L$, i.e., $N_e<N_L$. 
Consequently, in the applied field effective masses of the minority spin fermions are enhanced when compared to those of 
the spin-majority fermions, as discussed in main text.}
\label{lattice}
\end{center}
\end{figure}

One of the basic microscopic models of the correlated fermions for the lattice systems 
is the Hubbard model, accounting for a balance between the kinetic energy of moving 
fermions (hopping energy) and the effective local repulsive interaction $U$, whenever 
two fermions with opposite spins (or other internal quantum number) 
occupy the same site (cell) and the same orbital (Wannier) 
state (cf. Fig.\ref{lattice} for visualization of such state). 
Within this model, not only the principal predictions of the Landau Fermi-liquid 
theory have been derived\cite{vollhardt1984,vollhardt1987}, but also a unique 
explanation of the correlation-driven Mott metal-insulator transition\cite{gebhardmott}
and the emergence of d-wave pairing have been discussed\cite{scalapino2012}.    
Importance of the Hubbard model manifests itself also as the basis of derivation of the 
so-called $t-J$ model, which turned out to be very successful in 
description of the high temperature superconductivity in the cuprates (cf. e.g. Ref. [\cite{jedrak2011}]).

An important approach to solving the Hubbard model is the
variational Gutzwiller wave function (GWF)\cite{buenemann2012,kaczmarczyk2013} and its simplified version - the Gutzwiller approximation (GA)\cite{gutzwiller1965,gebhardmott}.
Those methods are based on the paradigm of optimizing the number of doubly occupied sites (cells) via which we optimize the local repulsive energy
balancing the renormalized by them the band energy. In effect, simpler GA approach leads among others to a renormalized mass of the quasiparticles 
in a direct manner which becomes divergent at the Mott-Hubbard localization \cite{bunemann2003,spalekj1990}, as well as leads concomitantly to
the infinite zero field magnetic susceptibility in the paramagnetic state\cite{brinkman1970}.
Although GA is strictly speaking, exact only for lattice of infinite dimensionality\cite{gebhardmott},
it may provide a good description of almost localized systems above the so-called upper critical dimensionality (not yet determined) for fermions.
Minimally, it plays the role of a mean field theory.     
Among other techniques frequently applied to the Hubbard model 
are also the dynamical mean-field theory (DMFT)\cite{parihari2011} or 
the quantum Monte Carlo (QMC)\cite{jarrell1992} methods.

Recently, we have discovered statistical inconsistencies 
in GA when one includes the magnetic field via the Zeeman term
and have provided the necessary correction to the 
GA approach. The extended approach called statistically-consistent Gutzwiller approximation (SGA)
was successfully also applied to the t-J\cite{jedrak2011,jedrak2010,spalek2011}, t-J-U\cite{abram2013} and to 
periodic Anderson models\cite{howczak2012} to describe both magnetism and superconductivity.
SGA is of the same class approach as GA (exact in
the infinite-dimension limit\cite{gebhardmott}), but additionally is consistent from the
statistical-physics point of view, as explained in detail below (cf. Appendix A).
 
The SGA method we are going to discuss here has its own merits. First of all, it is analytic and 
therefore can be applied to the infinite-size systems. 
This means also that it can be compared directly with more numerically oriented approaches, 
where the lattice size is usually limited. Second, it generalizes the Landau concept of the quasiparticle, 
as well as provides the corresponding effective single-particle  Hamiltonian. Third, it corrects the 
principal inconsistency of GA while retaining its attractive qualitative feature
by providing a testable approach  around the border between the moderate and the strong correlation limits.
This is the most difficult regime as then the kinetic and interaction energies are of comparable amplitudes.

In the present paper we have applied SGA method to almost half-filled band situation, where
the physics turns out to be non-trivially different from the half-filled case but still the almost-localized
character of the system\cite{spalek1987,kokowski1989,korbel1997,kaczmarczyk2009} is seen, as discussed below.
Such situation with $N_e<N_L$ (there are less fermions than cells) is physically feasible for neutral
fermions, whereas for the charged fermions (e.g., electrons) consideration of the non-half-filled band situation requires
an additional justification of preserving the charge neutrality of the whole system unless a compensating charge reservoir is assumed.
We discuss the results concerning magnetization, magnetic susceptibility, specific heat, 
as well as the spin-direction dependent effective masses of quasiparticles, in addition to metamagnetism.
As an example of concrete application we have also compared resulting from our approach magnetization results
with the experimentally obtained magnetization curve\cite{wiegers1991} for liquid \he\ and found a good semiquantitative 
agreement. The fitting parameter $U$ places this system in the moderate-correlation regime at ambient pressure.

The structure of the paper is as follows. In Sec. 2 we discuss the Hubbard model solution within the statistically consistent
 Gutzwiller approach (SGA). In Sec. 3 we provide main results 
concerning the dependence of the system properties on applied magnetic field. Sec. 4 contains conclusions and a brief overview. 
Appendix A provides some details of SGA.

A brief methodological remark is in place here. The description is applicable, strictly speaking, to neutral fermions as we include
applied magnetic field only via Zeeman term and ignore the Landau-level structure appearing for charged quantum particles. This approach may be then applied to the 
discussion of both liquid \he\ and to the neutral cold-atom systems of spin $\frac{1}{2}$ in optical lattice systems.

\section{Modeling: Statistically consistent Gutzwiller approach (SGA)}
Our starting point is the single-band Hubbard Hamiltonian in its translationally 
invariant form and in the applied magnetic field $H_a$
\begin{equation}
\mathcal{H}=\sum_{<ij>,\sigma}t_{ij}\hat c_{i\sigma}^{\dagger}\hat c_{j\sigma}+
U\sum_i\hat n_{i\uparrow}\hat n_{i\downarrow}-\sigma h\sum_{i,\sigma}\hat n_{i\sigma},
\end{equation}
where the first term expresses single-particle hopping between sites $i$ and $j$
(with the hopping amplitude $t_{ij}<0$) and the second describes the intra-atomic repulsive
interaction characterized by the Hubbard parameter $U>0$. The applied magnetic field 
is introduced via the Zeeman splitting (last term), with $h\equiv\frac{1}{2} g\mu_BH_a$, 
and for spin $\frac{1}{2}$ we have that $\sigma=\pm1$. In the case of neutral 
fermions (atoms in either liquid \he\ or in an optical lattice) magnetic field enters the Hamiltonian
 only via this term. Applying the variational GA procedure\cite{jedrak2010} we obtain the effective
one particle renormalized Hamiltonian in the form:
\begin{equation}
\label{ham}
 \mathcal{H}_{GA}= \sum_{{\bf k}\sigma}(q_\sigma(d,n,m)\epsilon({\bf k})-
\sigma h)\hat c_{{\bf k}\sigma}^\dagger\hat c_{{\bf k}\sigma}+LUd^2.
\end{equation}
where $\epsilon(\g{k})$ is the single-particle dispersion relation, 
$L\equiv N_L$ - the number of lattice sites, $d^2$ - the probability of having a doubly occupied 
site, $n= n_\sigma+n_{\bar{\sigma}}$ is the band filling,
and $m= n_\sigma-n_{\bar{\sigma}}$ is the uniform magnetic polarization per site. 
The band narrowing factor $q_\sigma(d,m,n)$ is derived from combinatorial 
calculations in the usual form (a transparent derivation is provided in Refs. [\cite{vollhardt1984}] and [\cite{gebhardmott}])
{\small
\begin{equation}
\label{qodsigma}
 q_\sigma=\frac{2\bigl[\sqrt{(n+\sigma m-2d^2)(1-n+d^2)}+
\sqrt{(n-\sigma m-2d^2)d^2}\bigr]^2}{(n+\sigma m)(2-n-\sigma m)}.
\end{equation}}
However, in an applied magnetic field the standard self-consistent GA procedure is insufficient. 
For achieving the statistical consistency one needs to introduce additional constraints. The essence of 
the SGA method  is discussed in Appendix A.
In effect, we introduce new effective Hamiltonian in the form
\begin{equation}
 \mathcal{H}_{SGA}\equiv \mathcal{H}_{GA}-\lambda_m(\sum_{i,\sigma}\sigma \hat n_{i,\sigma}-mL)-\lambda_n(\sum_{i,\sigma}\hat n_{i,\sigma} -nL),
\label{HSGA}
\end{equation}
where the Lagrange multipliers $\lambda_m$ and $\lambda_n$ represent the constraints imposed on $m$ and $n$ that
their values calculated self-consistently coincide with those determined variationally. 
The diagonalization of (\ref{HSGA}) allows for calculation of the thermodynamic potential functional
\begin{equation}
 \mathcal{F}^{SGA}=-\frac{1}{\beta} \sum_{{\bf k} \sigma}ln[1+e^{-
\beta E_{{\bf k} \sigma}^{(SGA)}}]+L(\lambda_nn+\lambda_mm+Ud^2),
\label{5}
\end{equation}
where the eigenenergies of quasiparticles are
\begin{equation}
 E_{{\bf k} \sigma}^{(SGA)}=q_\sigma\epsilon_{{\bf k}} -\sigma(h+\lambda_m) - \mu-\lambda_n.
\label{6}
\end{equation}
The functional (\ref{5}) represents the effective Landau functional with order parameters and
extra variables expressing the inter-particle correlations.

We see that in order to achieve that the polarization $m$ and the chemical potential $\mu$
determined variationally would coincide with those determined
in a self-consistent manner, we have effectively introduced corresponding {\it effective fields} 
adding to both h and $\mu$.  In essence, this procedure assures the fulfillment  
of the Bogoliubov theorem, discussed originally in the Hartree-Fock 
approximation, and stating that the introduced effective single-particle approach represents 
the optimal single-particle representation of the mean-field state.
Note also that the Luttinger theorem for Hamiltonian (\ref{HSGA}) is obeyed,
so that the system is represented  by a Fermi liquid, i.e., there is a one-to-one correspondence
 between the bare states (of energies $\epsilon_\g{k}$) and the 
quasiparticle states ($E_{\g{k},\sigma}$ for $h=0$). A number of renormalizing factors ($q_\sigma,\lambda_m,\lambda_n$)
appears though; those are determined by the statistical consistency (equilibrium) conditions.  The
presence of those three parameters determined either form self-consistency  conditions or variationally represent the features
which do not appear in the original Landau theory of Fermi liquid.This is the reason why it is termed either as a correlated or an
almost localized Fermi liquid.

The equilibrium values of parameters and mean field are obtained from the following minimizing
procedure of generalized grand-potential functional (\ref{5}) with respect to the variables assembled into a 
vector $\vec\lambda\equiv(m,d,n,\lambda_n,\lambda_m)$ and
representing all relevant quantities, determined from the necessary condition for minimum
\begin{equation}
\Big(\frac{\partial \mathcal{F}}{\partial \vec\lambda }\Big)_0=0,
\label{set}
\end{equation}
in combination with comparison of $\mathcal{F}$ values for different possible solutions.
Explicitly, the above conditions can be rewritten as a set of five self-consistent equations for the corresponding quantities
\begin{equation}
\left\{ 
\begin{array}{l}
  \displaystyle{\lambda_n=-\frac{1}{L}\sum_{{\bf k}\sigma}  \frac{\partial q_\sigma}{\partial n} f ( E_{{\bf k} \sigma}^{(SGA)} )\epsilon_{{\bf k}}},\\
  \displaystyle{\lambda_m=-\frac{1}{L}\sum_{{\bf k}\sigma}  \frac{\partial q_\sigma}{\partial m} f ( E_{{\bf k} \sigma}^{(SGA)} )\epsilon_{{\bf k}}},\\
  \displaystyle{d=-\frac{1}{2LU}\sum_{{\bf k}\sigma}  \frac{\partial q_\sigma}{\partial d} f ( E_{{\bf k} \sigma}^{(SGA)} )\epsilon_{{\bf k}}},\\
  \displaystyle{ n=\frac{1}{L}\sum_{{\bf k}\sigma} f ( E_{{\bf k} \sigma}^{(SGA)} )},\\
  \displaystyle{m=\frac{1}{L}\sum_{{\bf k}\sigma}  \sigma f ( E_{{\bf k} \sigma}^{(SGA)} )},
\end{array}
\right.
\label{eset}
\end{equation}
where $f(E)$ is Fermi-Dirac distribution function.
Note that the first two equations contain the derivatives of the band narrowing 
factor with respect to the respective variables and would be absent in the ordinary
Hartree-Fock approximation. In the case of GA, the fields $\lambda_m=\lambda_n\equiv0$; this is the limit 
of weak coupling and represents one of the checkouts on the method reliability.
The grand-potential functional evaluated for the optimal values of components of vector $\vec\lambda$,
and determined from (\ref{eset}) reduces to the physical grand potential $\Omega$.
Once we have determined the equilibrium thermodynamic potential,
we can also determine all relevant thermodynamic quantities. For example, the entropy is
\begin{equation}
\label{omega}
\begin{split}
 -S=\frac{d \Omega}{d T}
= \Big(\frac{\partial \mathcal{F}}{\partial T}\Big)_0 +
\Big(\frac{\partial \mathcal{F}}{\partial {\vec\lambda} }\Big)_0 \cdot\frac{\partial {\vec \lambda} }{\partial T},
\end{split}
\end{equation}
where subscript ``$0$'' labels the equilibrium values of variational parameters.
Since $\Big(\frac{\partial \mathcal{F}}{\partial \vec\lambda }\Big)_0=0$ 
from the (\ref{set}),
Eq. (\ref{omega}) simplifies to the form
\begin{equation}
\begin{split}
 S&=-\Big(\frac{\partial \mathcal{F}}{\partial T}\Big)_0=k_B\sum_{{\bf k}\sigma}\Big[ln(1+e^{-\beta E^0_{{\bf k}\sigma}})+\beta E^0_{{\bf k}\sigma}
f( E_{{\bf k} \sigma}^{0} )\Big].
\end{split}
\end{equation}
The specific heat is then defined in the usual manner
\begin{equation}
 c_V= T\frac{dS}{dT}\Big|_{n,T,h,V}\equiv-T\frac{\partial^2 \mathcal{F}}{\partial T^2}\Big|_0.
\label{spec}
\end{equation}
Note that variational parameters depend on temperature in a non-trivial manner. 
Therefore we have to determine the specific heat numerically. A detailed analysis of the results follows next.

\section{Results and discussion}
\subsection{Principal physical properties}
To obtain the equilibrium values from solutions of the set of equations (\ref{eset}), 
we have made use of the scientific library GSL on the GNU license\cite{GSL}; 
the precision of the numerical results was $10^{-7}$. In our calculations 
we set energy scale in the units of nearest neighbors hopping $|t|$. 
We consider a particular closed-packed structure, the 
three-dimensional face-centered cubic lattice of size 200x200x200. 
If not specified otherwise, we have taken into account the 
second-nearest-neighbor hopping set as $t'=0.25$, the band filling as $n=0.97$, and the reduced temperature 
in the system as $\beta\equiv\frac{1}{k_BT}=500$, which can be regarded practically as $T\approx0$ limit. 

\begin{figure}[t]
\begin{center}
\includegraphics[width=90mm]{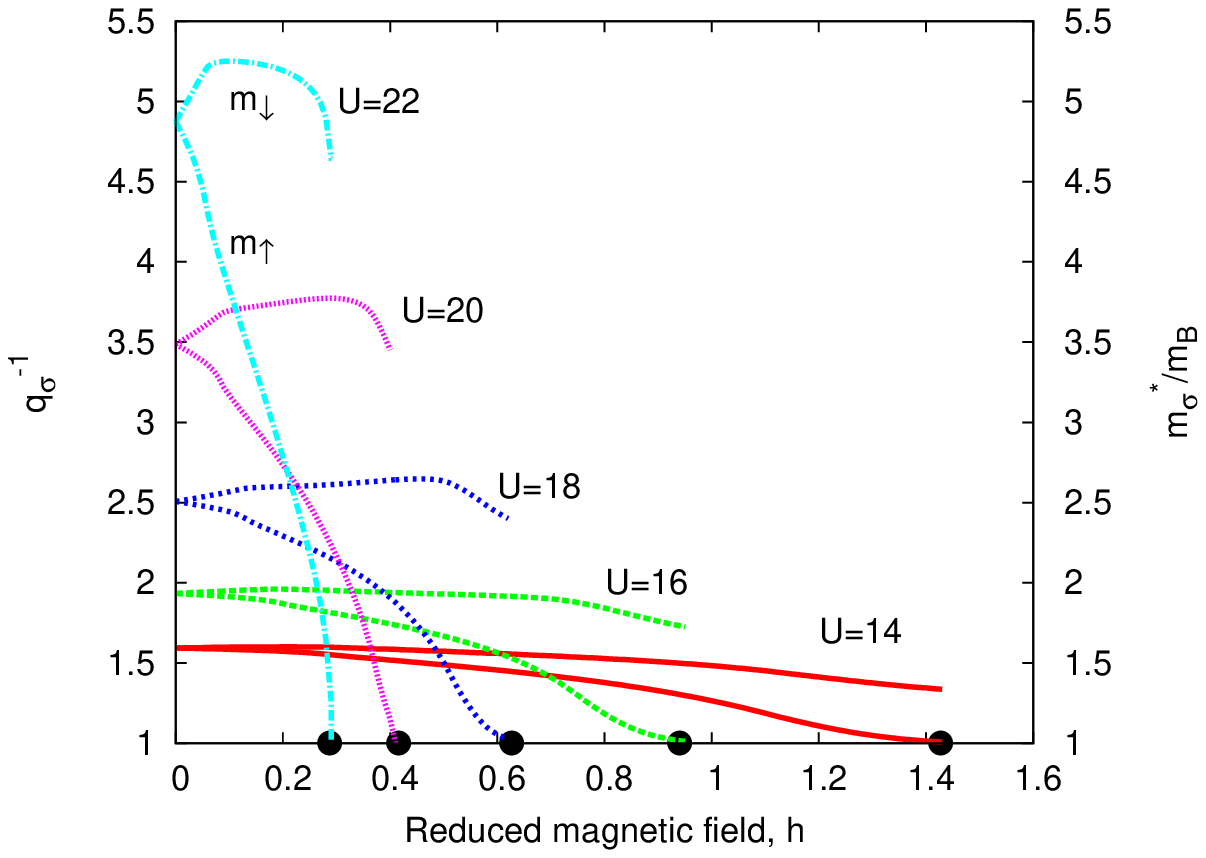}
\caption{Spin-direction dependent effective quasiparticles masses 
as a function of applied magnetic field for an almost half-filled band 
($n=0.97$). The curves terminate when the magnetization saturates, 
i.e., when in the spin-majority subband the quasiparticle mass 
$m_\uparrow=m_B$ (the solid points on the horizontal axis). The evolution 
is spin asymmetric since $q_\sigma$ depends on $m$ in this fashion with respect to
spin direction $\sigma=\pm1$.}
\label{qs}
\end{center}
\end{figure}

Both approaches, GA as well as SGA, automatically account for renormalization of effective mass
($m^*_\sigma$) with respect to the bare band mass ($m_B$) in the form:
\begin{equation}
 \frac{m^*_\sigma}{m_B}=q_\sigma^{-1}\equiv(\frac{\partial E_\g{k}}{\partial \epsilon_\g{k}})^{-1},
\end{equation}
where $q_\sigma$ is the optimized Gutzwiller band-narrowing 
factor defined in (\ref{qodsigma}). In the absence of applied magnetic field 
the effective mass for larger values of the interaction is renormalized appreciably, 
as detailed in Fig.\ref{qs}. However, in the applied magnetic field, in an 
almost half-filled band we observe their strongly asymmetric dependences 
with respect to the spin direction.
This asymmetry with respect to the value of $\sigma=\pm1$ is caused by the corresponding 
$m$ dependence of $q_\sigma$. The curves terminate at the 
saturation point $m=n$, where simultaneously $m_\uparrow=m_B$ and the spin-minority 
quasiparticle subband becomes empty. Parenthetically, 
the saturation point may be used thus to determine the value of bare band mass, $m_B$, as the 
Hubbard interaction is then switched off. 
This spin-dependent mass renormalization of quasiparticles in the strong correlation regime
has been discussed extensively in the literature \cite{kaczmarczyk2009,maska2010,bauer2007,bauer2009}, 
and observed experimentally by means of the de Haas-van Alphen oscillations in strong magnetic 
fields\cite{flouquet2005,sheikin2003}. It thus represents a crucial new concept, which does not appear in the 
standard Landau Fermi liquid (LFL) theory.

\begin{figure}[t]
\begin{center}
\includegraphics[width=90mm]{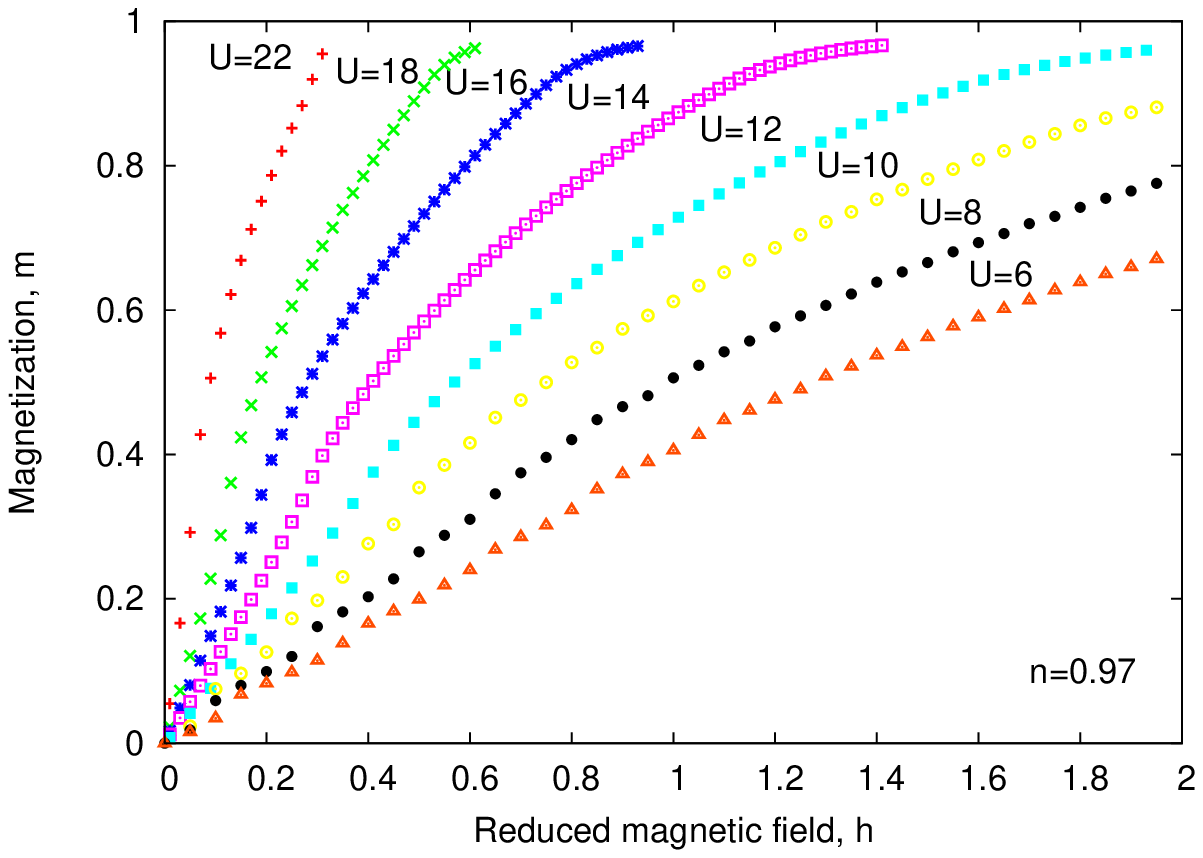}
\caption{Magnetization curve for selected values of Hubbard parameter $U$ and fixed band filling $n=0.97$. 
Even for very large values of $U$ the curves show a 
metamagnetic-like transition in the form of a kink. }
\label{mag}
\end{center}
\end{figure}

\begin{figure}[t]
\begin{center}
\includegraphics[width=90mm]{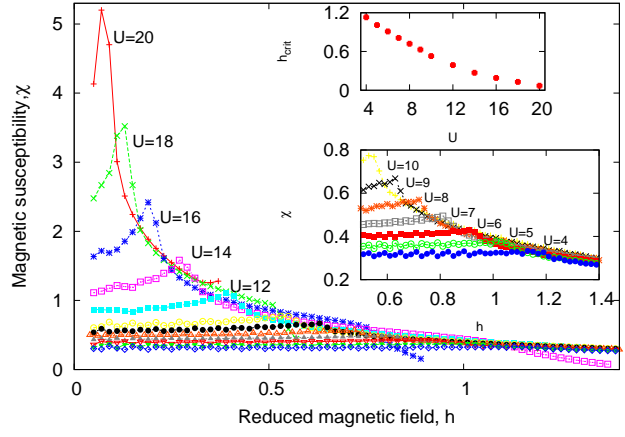}
\caption{Magnetic susceptibility ($\chi=\frac{\partial m}{\partial h}$) as a function of reduced magnetic field for selected
 values of the Hubbard parameter $U$. 
In the whole range of $U$ values we observe a discontinuity which
we ascribe to the metamagnetic-like behavior. In the lower inset we plot the zoomed
curves for smaller values of $U$ parameter. In the upper inset we present dependence 
the critical field ($h_{crit}$) representing the metamagnetic-like point for selected
values of parameter $U$.}\label{sus}
\end{center}
\end{figure}

\begin{figure}[t]
\begin{center}
\includegraphics[width=90mm]{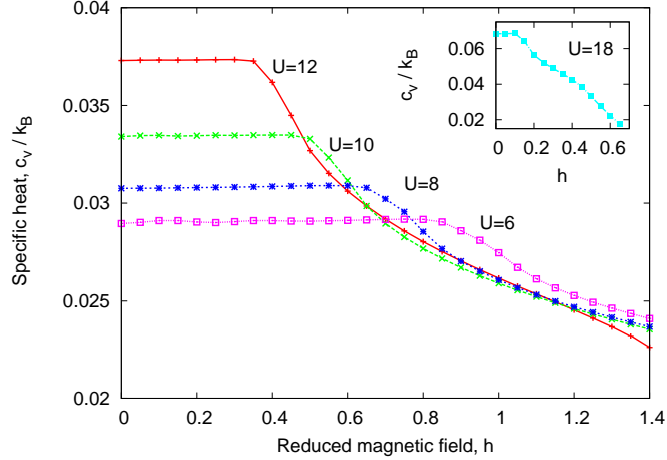}
\caption{Specific heat (per site) as a function of reduced magnetic field for selected values of
$U$. The slope change occurs for the same values of $h$ 
as the kinks in the magnetic susceptibility shown in Fig.\ref{sus}. 
At lower fields the mass enhancements $\{m_\sigma^*\}$ are symmetric, 
so their contributions to the total density of states average out 
to the value roughly independent of $h$. The temperature was taken as $k_BT=0.04|t|$.}\label{cvh}
\end{center}
\end{figure}

In Fig.\ref{mag} we show a family of magnetization curves as a function of 
reduced magnetic field ($h\equiv\mu_BH_a$). In comparison to the results obtained 
within GA\cite{vollhardt1984,korbel1995} we do not observe any spectacular metamagnetic transition for $n<1$. 
The magnetization curves, especially in the intermediate interaction 
regime, are rather smooth and saturate gradually. Nonetheless we observe a weak kink which we qualify as 
a metamagnetic-like behavior.
Indeed this kink can be singled out clearly on the field dependence of the  magnetic susceptibility and the specific heat 
curves, as shown in Figs.\ref{sus} and \ref{cvh}, respectively.

One can say that the presence of both spin-dependent masses and the metamagnetic behavior signal an appearance of the so-called 
almost-localized Fermi liquid (ALFL) state, as such behavior is absent for the Landau Fermi liquid \cite{korbel1995}. 
The diminution of $c_V$ with $h$ is the sign of a combined decrease of both the effective-masses components, as well as 
of the decreasing population of the spin-minority subband.

\subsection{Subsidiary quantities}

Introduction of constraints results in additional self-consistent equations to be obeyed, 
what in turn restricts the parameter space to look for the proper global minimum
of the free-energy functional. Furthermore, the Lagrange multipliers  $\lambda_m$ and
$\lambda_n$ play the role of the effective field and the shift of the chemical potential, respectively.
They are induced by the correlations in the sense that their presence is induced by the 
presence of $q_\sigma$ and particularly, by its singular dependence on both $m$ and $n$ in the present situation.
The fields $\lambda_m$ and $\lambda_n$ are determined within the variational/self-consistent procedure
and represent the third factor distinguishing ALFL from LFL. For the sake of completeness, we provide in Table
\ref{coef} the exemplary values of the relevant quantities. Note that $\lambda_n$ almost compensates $\mu$ and
in effect we obtain the effective value of the Fermi energy for the heavy quasiparticles.

\begin{table}
\begin{center}
\caption{ Values of the parameters obtained for $U=12$, $t'=0.25$, $n=0.97$, $\beta=500$ (in units where $|t|=1$) 
for four values of reduced magnetic
field. The calculations were made for the fcc lattice 200x200x200.\vspace{2mm}}
\label{coef} {
\begin{tabular}{|c|c|c|c|c|}
\hline
 Quantity 	 & {$h=0$}    & {$h=0.2$}  & {$h=0.5$}&{$h=1$}\\   \hline
$\mathcal{F}$	 & -6.4752954 & -6.4765977 & -6.4354871&-6.0518362 \\ 
$d$		 & 0.3231845  & 0.31933441 & 0.2950449& 0.2370746\\ 
$m$		 & 0. 	      & 0.1690908  & 0.4483957&  0.7206020\\  
$\mu$ 		 & 5.8387352  & 5.8227839  & 5.6845669& 4.9815074\\  
$\lambda_n$	 & -5.5747470 & -5.5601127 &-5.4599753& -5.2141073\\ 
$\lambda_m$	 & 0. 	      & 0.4007685  &1.1410240& 2.1809003\\
$q_\uparrow$	 & 0.7250737  & 0.7293968  &0.7482932& 0.7971271\\
$q_\downarrow$    & 0.7250737  & 0.7236794  &0.7307299& 0.7569461\\\hline
\end{tabular}
}
\end{center}
\end{table}

\begin{figure}[t]
\begin{center}
\includegraphics[width=90mm]{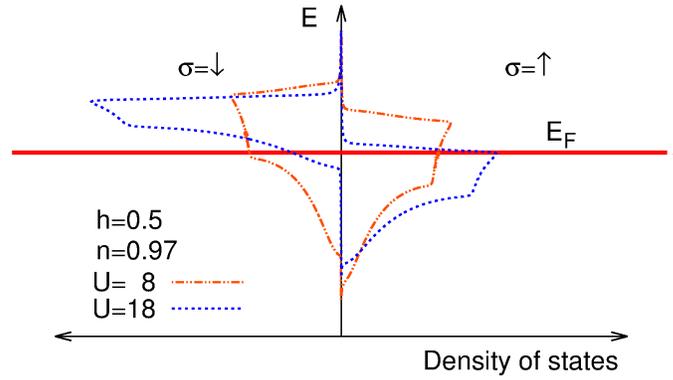}
\caption{Spin resolved density of states of quasiparticle states in the applied magnetic field $h=0.5$ for the selected values of $U$.
The spin (magnetic-moment) minority subbands is narrowed down, whereas the spin-majority-spin subband 
widens up so that it acquires the bare bandwidth as the saturation state is reached in the strong-field limit. }\label{DOS}
\end{center}
\end{figure}

\begin{figure}[t]
\begin{center}
\includegraphics[width=90mm]{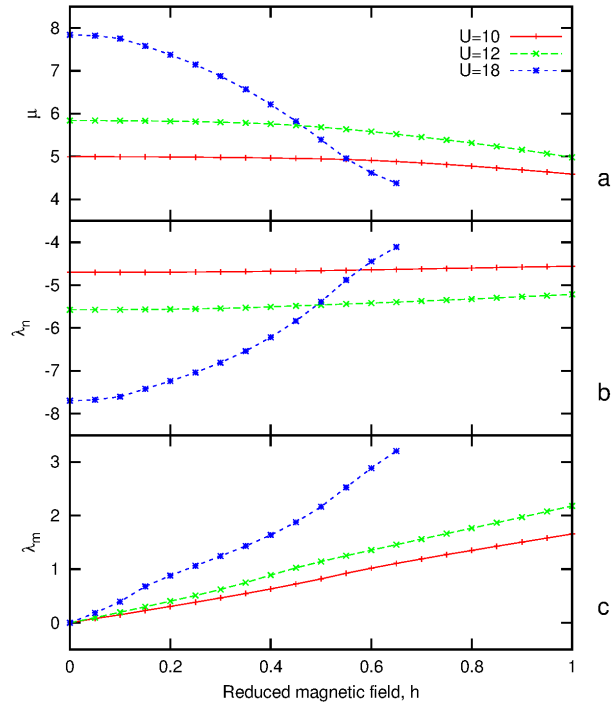}
\caption{Panels a-c show evolution of fields $\lambda_m$ and $\lambda_n$ and chemical potential $\mu$ with respect to the applied magnetic field.
The field $\lambda_m$ appears only in the spin polarized state, whereas $\lambda_n$ almost compensated $\mu$.
}\label{panels}
\end{center}
\end{figure}

\begin{figure}[t]
\begin{center}
\includegraphics[width=90mm]{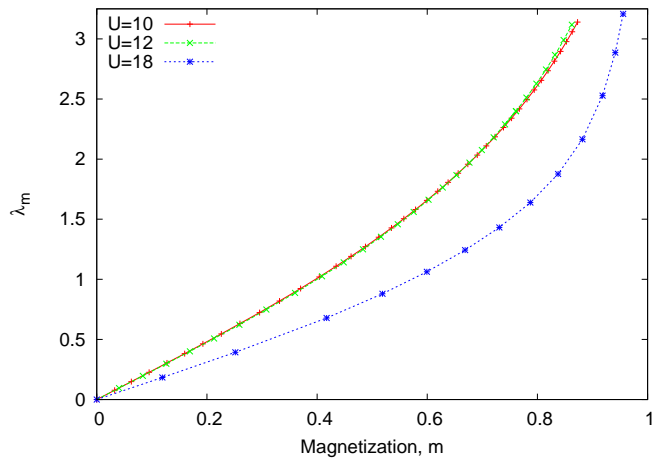}
\caption{Effective field induced by correlations as a function of magnetization; it illustrates its nonlinear character and in this manner cannot be regarded as a Weiss molecular field.}\label{mol}
\end{center}
\end{figure}

In Fig. \ref{DOS} we plot the spin-resolved density of states for the two values of 
$U=8$ and $18$ (in units of $|t|$) and in an applied field. The spin subbands are shifted by the amount $h+\lambda_m$ and narrowed down asymmetrically by the spin-direction
dependent factor $q_\sigma$.
In the panel displayed as Fig.\ref{panels}a-c we have plotted  $\lambda_m$, $\lambda_n$,
and $\mu$, all as a function of $h$. Additionally, the field $\lambda_m$ as a function of magnetization $m$ is shown in Fig. \ref{mol}. 
This effective field is a nonlinear function of both $h$ and $m$ and represents a relatively
fast growing quantity with the increasing $h$. It is only weakly $h$ dependent in the moderate interaction limit. 
This means that the presence
of  $\lambda_m$ will have an essential impact of magnetic properties, while a relative constancy of $\mu$
and $\lambda_n$ in the regime $h<h_{crit}$ explains a flat behavior of $c_V(T)$ in that regime. It would be interesting to see if 
the presence of the effective field $\lambda_m$ acting on the spin degrees of freedom only (in analogy to the Weiss molecular field) 
can be detected with the same accuracy, as the spin splitting of the masses. Such test would be a decisive step forward in defining
an almost localized Fermi liquid as a separate state from the Landau Fermi liquid. Note however, that $\lambda_m$ cannot be, strictly 
speaking, considered as a molecular Weiss field, since it is nonlinear in $m$.

\subsection{Concrete example: liquid \he}
\begin{figure}[t]
\begin{center}
\includegraphics[width=90mm]{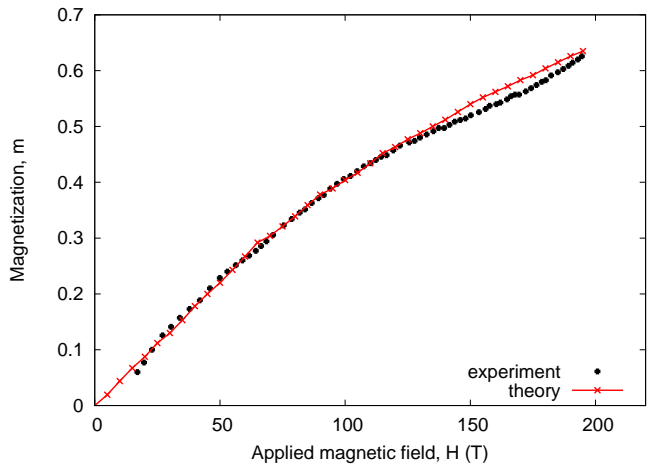}
\caption{Magnetization in dimensionless units, $m\equiv n_\uparrow n_\downarrow$ vs. applied field $H$ 
calculated within SGA fitted to the experimental 
data \cite{wiegers1991}, Fitting parameters are: $U=5|t|$, $n=0.95$, $t'=0.25|t|$, $|t|=62.5 K$.
A very weak, almost unnoticeable metamagnetic kink is present. The reason for the small deviation 
of the results from the data (for $H\gtrsim 125T$) is not accounted for
within the present approach.}
\label{he}
\end{center}
\end{figure}

Liquid \he\ is regarded as a canonical example of the Landau Fermi liquid (cf. e.g. Ref. \cite{dobbs}). It has
also been regarded as an ALFL, as it undergoes a transition to the solid state at the pressure  $p\simeq 34\ bar$,
with the localization of the \he\ atoms regarded as a fermions of spin $\frac{1}{2}$\ {}\ \cite{vollhardt1984,vollhardt1987,spalek2000}.
Here we discuss briefly its behavior in an applied magnetic field at ambient pressure.
We have compared magnetization curve obtained from our model to the experimental magnetization\cite{wiegers1991} 
of the liquid \he, as shown in Fig. \ref{he}. We find a good 
overall agreement. Note that our results, obtained for $n<1$ do not exhibit any discontinuous 
metamagnetic transition which disqualified the applicability of the standard GA to \he\ and with $n=1$ \cite{vollhardt1984}.
To avoid this discontinuity,
we have introduced about 5\% of vacancies in this virtual fcc
lattice representing liquid \he. This number of vacant sites provides the best fit of our results to the experimental data.
Furthermore it is important to note that the relatively small number of holes introduced preserves
the almost localized nature of this quantum liquid.  This note is important also in the view of the circumstance, 
that the detailed $m(h)$ dependence is very sensitive to the number of quantum vacancies.
It is tempting to suggest that the effective empty-site content
$\delta\equiv 1-n$ of the order of few percents can be interpreted as a presence of quantum 
vacancies in the liquid state, in direct analogy to the quantum Andreev vacancies postulated and observed  in solid \he\cite{andreevnew}.
From a detailed analysis of the fitting procedure we conclude that the assumption of a
non-half-filled lattice is necessary if either GA, SGA or DMFT approaches and the Hubbard
model are to emulate the magnetic behavior of liquid \he.

One important difficulty of the just discussed results (Fig. \ref{he})) should be raised at this point. Namely,
a relatively small value of $U\sim5|t|\ll W$ (where $W\equiv24|t|$ is bare band width) 
obtained from the $m(h)$ best fit places the liquid \he\ at ambient pressure as a moderately correlated system.
This seemingly unexpected conclusion is also 
in agreement with the DMFT analysis\cite{parihari2011}. In this way, the Hubbard-model approach squares also 
qualitatively with the fundamental Fermi-liquid theory assumption that the interaction between quasiparticles can be essentially
limited to the region very close to the Fermi surface ($U$ substantially smaller than $W$). Furthermore, 
the method for estimating the number of 
vacancies might be associated with a detailed study of the field dependence of the specific heat to determine a
possible spin-direction dependence of the effective mass, though the weak effect for this moderate value of $U$ may be prove decisive.
 Such an explicit spin-direction mass dependence 
is absent when we have exactly half-filled situation (i.e., one atom per site, without any vacant cells).
 
Additionally, in spite of this success in explaining $m(h)$ in a semiquantitative manner,
 there still is another problem concerning
determination with the same set parameters of a realistic value of the mass enhancement estimated from the $c_V(T)$ 
curves for $T\rightarrow0$\cite{greywall1983,greywall1986}. Namely, the value of the specific-heat enhancement at $h=0$ 
obtained for $U=5|t|\sim W/5$, is far too small to 
provide the corresponding effective mass enhancement $m^*=2.86 m_0$ at ambient pressure, where $m_0$ is the 
free \he\ atomic mass. This disagreement is common to the present and the previous\cite{vollhardt1987,parihari2011} treatments.
A small mass enhancement within GA, SGA or DMFT of the mass can be interpreted in two ways. First, the spin-fluctuation contribution can be
very important\cite{li1990}.
In that situation, SGA should be considered only as a saddle-point approximation to a more complete approach (cf. Appendix B).
 This question 
certainly requires a detailed analysis. 
Second, one has to note, that the $m^*/m_0$ ratio estimate from the
 experimental $c_V$ data\cite{greywall1983,greywall1986}, based on the relations to an ideal gas, may not be fully adequate either. 
Finally, our assumption that $N_e<N_L$ and $\delta\equiv1-n$ independent of $h$ or pressure, may be analyzed further as well.

\section{Conclusions}
We have carried out a systematic study of a correlated Fermi liquid modeled by the Hubbard model on fcc lattice, 
utilizing the statistically consistent Gutzwiller approach (SGA)\cite{reprint}
in an applied magnetic field and in an almost half-filled regime. 
Within this method, the field and spin direction dependent effective masses, 
the magnetization curve, the magnetic susceptibility, and the specific heat, 
were all calculated in a both self-consistent and variationally optimal 
 manner. We have found a metamagnetic-like behavior
 tracing the discontinuities which would appear for the half-filled band case (one particle per cell).
In other words, we have not obtained any first-order metamagnetic transition which limits the original Gutzwiller approach (GA) 
applicability to real systems such as liquid \he\ (see also the discussion in Ref. \cite{parihari2011}).
Apart from that, GA approach is not statistically consistent, as discussed in detail in the text.
We have found a good overall agreement of our approach with experimental data for the 
liquid \he\ magnetization. However, there remains a question of putting into a mutual agreement 
the magnetization data with the magnitude of effective-mass enhancement observed for that system in zero applied field.
The quantitative analysis within the lattice-approach required introducing a small number ($\sim5\%$) of vacant cells
to destroy the strong metamagnetism. A way of including the spin fluctuations is suggested (cf. Appendix B), 
starting from the SGA state, which
should be then regarded as a saddle-point state for a more complete analysis.
In this manner, SGA state replaces the Hartree-Fock state regarded so far as a reference (saddle-point) state for further
considerations containing spatially inhomogeneous fluctuations\cite{hertz1976,moriya1985, millis1993}, here of both spin and charge types.

In summary, our SGA approach, as well as the works carried within DMFT \cite{parihari2011, bauer2007, bauer2009} provide a convergent quasiparticle 
language in the sense that effective masses are spin-direction dependent in the magnetically polarized state. We also introduce a set of nonlinear 
effective fields determined in a self-consistent manner which renormalize both the chemical potential and the applied field magnitude experienced by the 
quasiparticles representing the almost localized Fermi liquid. Our method of approach may be regarded as equivalent to the slave-boson approach 
in the saddle-point approximation\cite{reprint}, without involving slave-boson fields which introduce spurious Bose-condensation phase transitions at nonzero temperature.

\section*{Acknowledgements:} 
We would like to thank Marcin Abram, Jakub J\c{e}drak, and Jan Kaczmarczyk for very useful discussions, and technical help.
We also greatly acknowledge cooperation with professor W\l odzimierz W\'ojcik on the early stages of this work.
The work has been partially supported by the Foundation for 
Polish Science (FNP) under the TEAM program, as well as by the 
National Science Centre (NCN) under the MAESTRO program, Grant No. DEC-2012/04/A/ST3/00342.
Preliminary results of this work have been presented in 
proceedings of XVI Training Course in the Physics of Strongly Correlated Systems
\cite{AIP}.  
\appendix
\section{Statistically consistent Gutzwiller approximation}

As pointed out in the main text, it has been noticed (see Ref. \cite{reprint,AIP,jedrak} and references therein)
that the Gutzwiller approximation which leads to the effective single-particle Hamiltonian (\ref{ham})
has a principal drawback. Namely if we try to calculate the ground state energy from $ E_G\equiv\langle\mathcal{H}_{GA}\rangle$
we have to determine first the average spin polarization $m$, the double occupancy probability $d^2$, and $\mu$ first\cite{acta}. 
This can be achieved in two ways: either by minimizing the free-energy functional $\mathcal{F}$ for $T\geq0$ (defined by Eq. (\ref{5} for $\lambda_m=\lambda_n=0$))
with respect to variables ($m,d^2,n$) or by writing down the self-consistent equation of $m$ and $n$ and minimizing $\mathcal{F}$ with respect to $d^2$.
Those two procedures should provide the same answers if (\ref{ham}) represents properly defined single-particle Hamiltonian.
In fact, the results do differ\cite{reprint,jedrak} and this discrepancy is caused by the fact that the single-particle energies depend on $q_\sigma$ which depends
on $m$ and $n$ in a nonanalytic manner (see e.g. the limits $n\rightarrow1$ and/or $m\rightarrow1$). The presence of this discrepancy means that GA violates the
fundamental Bogoliubov theorem that the effective  single-particle states represents the optimal quasiparticle states in the variational sense,
with the ordinary self-consistent procedure preserved at the same time.

To cure this principal defect one can choose the approach in which one imposes the constraints so that the consistency form the point of view of
statistical mechanics is preserved. To carry through such a procedure one can utilize the maximum entropy method to derive the correct statistical
distribution (with constraints) and then proceed further\cite{jedrak}. Here we have chosen a slightly different path. Namely, we have defined the generalized
free energy functional (\ref{5}) with the constraints that the variationally calculated quantities $m$ and $n$ represent correct values.
This is realized by adding the two terms to $\mathcal{H}_{GA}$, as written in Eq. (\ref{HSGA}). This last step amounts to introducing statistically-consistent Gutzwiller
approximation (SGA). In result, we have to minimize $\mathcal{F}$ given by (\ref{5}) additionally with respect to $\lambda_m$ and $\lambda_n$ and the procedure introduces
two physically relevant fields (cf. (\ref{6})) which appear only in correlated state ($q_\sigma<1$) and vanish in the small-$U$ limit, i.e., in the Hartree-Fock limit
 ($q_\sigma\rightarrow1$). The last limiting situation represents thus a test for the correctness of our approach.

One should note that SGA is based  essentially on GA, but provides additionally essential consistency of the model by introducing two additional parameters into the
approach. In summary, the 
self-consistent equations
\begin{equation}
\left\{ 
\begin{array}{l}
  \displaystyle{\frac{1}{L} \sum_{ i,\sigma} \sigma \langle\hat n_{i\sigma} \rangle}= \displaystyle{\frac{1}{L}\sum_{{\bf k},\sigma}\sigma\langle \hat n_{\g{k}\sigma}\rangle\equiv m},\\
  \displaystyle{\frac{1}{L} \sum_{{ i},\sigma} \langle \hat n_{i\sigma}\rangle} = \displaystyle{\frac{1}{L}\sum_{{\bf k},\sigma}\langle \hat n_{\g{k}\sigma}\rangle\equiv n},\\
  \displaystyle{\langle \hat n_{\g{k}\sigma}\rangle} = \displaystyle{\frac{1}{e^{\beta(E_{\g{k}\sigma}-\mu)}+1}},
\end{array}
\right.
\end{equation}
are consistent with variational equations (\ref{set}) derived within the SGA method.
\section{SGA as a saddle-point approximation}
Here we would like to sketch the method of extending the SGA approach. In this method SGA is regarded as a saddle-point approximation to a more complete 
theory which includes the quantum fluctuations.

In general, the constraints in (\ref{HSGA}) should be written in th local form
\begin{equation}
\begin{gathered}
 -\bigg{\{} \sum_{i\sigma} \lambda_{m i}\sigma(\hat n_{i\sigma}-\langle \hat n_{i\sigma}\rangle)+\sum_{i\sigma} \lambda_{n i}(\hat n_{i\sigma}-\langle \hat n_{i\sigma}\rangle)\bigg{\}}\\
\equiv-\bigg{\{}\sum_{i\sigma}( \lambda_{m i}\delta\hat S^z_i-\lambda_{n i}\delta\hat n_i)  \bigg{\}},
\end{gathered}
\end{equation}
with $\delta \hat S_i^z\equiv \hat S_i^z- \langle \hat S_i^z\rangle$ and $\delta \hat n_i\equiv\hat n_i-\langle \hat n_i\rangle$.
In other words, in the system with fluctuating local quantities, the constraints should be obeyed also locally. Also, the charge fluctuations are expressed via the term
\begin{equation}
 U\sum_i( \hat n_{i\uparrow} \hat n_{i\downarrow}-\langle \hat n_{i\uparrow}\hat n_{i\downarrow}\rangle).
\end{equation}

The last term can be rewritten in the spin explicit form
\begin{equation}
\begin{gathered}
 \hat n_{i\uparrow} \hat n_{i\downarrow}=\bigg{(}\frac{n_{i\uparrow} -\hat n_{i\downarrow}}{2}\bigg{)}^2+\frac{\hat n_{i}^2}{4}\\
= (\delta \hat S^z_i)^2+2\langle\hat S^z_i\rangle\delta \hat S^z_i+\langle\hat S^z_i\rangle^2 +\frac{\hat n_{i}^2}{4},
\end{gathered}
\end{equation}
where $\langle\hat S^z_i\rangle\equiv m$ and $\hat n_i\equiv \hat n_{i\downarrow}+\hat n_{i\uparrow}$ . In effect, we have 
\begin{equation}
 U\sum_{i}\hat n_{i\uparrow} \hat n_{i\downarrow}=U\sum_i\delta (\hat S^z_i)^2+U\sum_i \frac{\hat n_i^2}{4} +const.
\end{equation}
One can reformulate the whole approach in the spin rotationally invariant form in a straight forward manner.
In effect, we have linear and quadratic spin and charge fluctuations.
We thus can define the part of Hamiltonian responsible for fluctuations as 
\begin{equation}
 \mathcal{H}=\mathcal{H}^{SGA}+\mathcal{H}^{fluct} +const,
\end{equation}
where
\begin{equation}
 \mathcal{H}^{fluct}\equiv-\sum_i\bigg{[}\lambda_{m i} \hat{\delta S_i^z}- \lambda_{n i}\delta\hat n_i\bigg{]} +U\sum_i\bigg{[} (\hat{\delta S_i^z})^2+\frac{1}{4}(\delta \hat n_i)^2\bigg{]} .
\label{B6}
 \end{equation}
This expression (\ref{B6})(or their spin-rotation invariant correspondants) should be inserted into the expression for the energy expressed in the functional-integral form
over the Fermi fields\cite{acta} and then evaluated explicitly, as it involves cumbersome calculations. Methodologically, it is analogous to the slave-boson approach\cite{li1990}.
However, the present approach involves only coherent physical Fermi $\{\hat c_{i\sigma},\hat c_{i\sigma}^\dagger\}$ and Bose $\{\lambda_m,\lambda_n\}$ fields, without introducing condensing 
ghost slave-boson fields.

When the reference (saddle-point) system is represented by an almost localized Fermi liquid, both the spin and the charge fluctuations become thus relevant. This should lead to
a renormalization of the thermodynamic properties and should be analyzed separately, as in the present paper we concentrate on the $T\rightarrow0$ results.
Nonetheless, inclusion of the fluctuations should lead to a further enhancement of the specific heat, as well as to the presence of the $T^3\ln{T}$ term for $h=0$\cite{li1990}.

\section*{Bibliography}
\bibliographystyle{cmpj.bst}

\providecommand{\noopsort}[1]{}\providecommand{\singleletter}[1]{#1}%

\end{document}